\documentclass[lettersize,journal]{IEEEtran}
\usepackage{amsmath,amsfonts}
\usepackage{algorithmic}
\usepackage{algorithm}
\usepackage{array}
\usepackage[caption=false,font=normalsize,labelfont=rm,textfont=rm]{subfig}
\usepackage{textcomp}
\usepackage{stfloats}
\usepackage{url}
\usepackage{verbatim}
\usepackage{graphicx}
\usepackage{cite}
\usepackage{makecell}
\begin{document}
\title{\textit{V2Sim}: An Open-Source Microscopic V2G Simulation Platform in Urban Power and Transportation Network}
\author{ Tao Qian,~\IEEEmembership{Member,~IEEE}, Mingyu Fang,~\IEEEmembership{Student Member,~IEEE}, Qinran Hu,~\IEEEmembership{Senior Member,~IEEE}, Chengcheng Shao,~\IEEEmembership{Member,~IEEE}, Junyi Zheng
        % <-this % stops a space
\thanks{Tao Qian, Mingyu Fang and Qinran Hu are with the School of Electrical Engineering, Southeast University, Nanjing 210000, China (e-mail: taylorqian@seu.edu.cn; mingyu.fang@seu.edu.cn; qhu@seu.edu.cn;). }% <-this % stops a space
\thanks{C. Shao is with the School of Electrical Engineering and State Key Laboratory on Electrical Insulation and Power Equipment, Xi’an Jiaotong University, Xi’an 710049, China (e-mail:c.c.shao22@msn.cn).}
\thanks{J. Zheng is with Star Charge Technology Co., Ltd, Changzhou 213000, China (e-mail: junyi.zheng@wbstar.com).}
\thanks{This platform is open-sourced under BSD 3 license on Github. Link: https://github.com/fmy-xfk/v2sim}}


\maketitle

\begin{abstract}
This paper proposes V2Sim, an open source Python-based simulation platform designed for advanced vehicle-to-grid (V2G) analysis in coupled urban power and transportation networks. By integrating a microscopic urban transportation network (MUTN) with a power distribution network (PDN), V2Sim enables precise modeling of electric vehicle charging loads (EVCL) and dynamic V2G operations. The platform uniquely combines SUMO for MUTN simulations and an optimized DistFlow model for PDN analysis, with dedicated models for fast charging stations (FCS) and slow charging stations (SCS), capturing detailed charging dynamics often overlooked in existing simulation tools. V2Sim supports a range of customizable V2G strategies, advanced fault-sensing in charging stations, and parallel simulation through multi-processing to accelerate large-scale case studies. Case studies using a real-world MUTN from Nanjing, China, demonstrate V2Sim’s capability to analyze the spatial-temporal distribution of EVCL and evaluate V2G impacts, such as fault dissemination and pricing variations, in unprecedented detail. Unlike traditional equilibrium models, V2Sim captures single-vehicle behavior and charging interactions at the microscopic level, offering unparalleled accuracy in assessing the operational and planning needs of V2G-compatible systems. This platform serves as a comprehensive tool for researchers and urban planners aiming to optimize integrated power and transportation networks.
\end{abstract}


\section{Introduction}

\IEEEPARstart {T}{he} power distribution network (PDN) is being influenced more by the electric vehicle charging load (EVCL) due to the rise in the number of electric vehicles (EVs) \cite{evcl_impact}. Since EVs are participants of the urban transportation network (UTN), the driving and charging patterns of their users have an impact on the distribution of EVCL in both space and time\cite{st_impact}. As a result, it is important to conduct a survey of the EVCL generating mechanism and distribution, which may play an auxiliary role in planning EV charging infrastructure (EVCI) and deploy EV charging stations (EVCSs).

Early researches surveyed EVCL by user equilibrium (UE) models. The Wardrop UE model \cite{UE} and the stochastic user equilibrium (SUE) \cite{SUE} model are the earliest and long-tested models for solving traffic assignment problem (TAP) whose goal is to determine the traffic flow of links by given OD pairs. By adding EVCSs into the UTN and considering EVCS constraints, UE models are able to solve the TAP in UTN with EVCS, and then calculate the flow and EVCL at each EVCS. An example is \cite{UE_EV1}, which discussed UE with battery EVs and the recharging of EVs. UE with EVCS planning is referred in \cite{UE_EV2}, which utilized a bi-level model to allocate the facilities and their capacity in the upper level and characterize UE behavior in the lower level. 

To model EVCL, taking both UTN and PDN into consideration simultaneously is also a common option. Coupled UTN and PDN were referred in \cite{couple}, and such coupled optimization is also able to consider the constraints of fast charging stations (FCSs), PDN, charging service time, mixed EVs and diesel vehicles, and emission \cite{couple, mixed, bagli, emit}. EVCL could be predicted on the basis of UE \cite{EVCL_UE}, and coupled simulations of UE and EVCL revealed the impact of FCS planning on traffic flow \cite{FCS_impact}.

Monte Carlo (MC) method is also frequently used in EVCL modeling and ECVI planning. Probability model of driving laws and charging characteristics for EVCL in considered in \cite{MC1}. A threshold of failed recharge attempts is added to the MC simulation in \cite{MC2}. MC method can also be utilized for EVCS analysis \cite{MC3}. A combination of ME method neural networks are also reported. In both \cite{MCNN1} and \cite{MCNN2}, EVCL is calculated by MC simulation and a neural network is utilized for EVCL prediction.
\IEEEpubidadjcol
The National Renewable Energy Laboratory (NREL) developed OpenPATH App for individuals to measure the energy consumption on their trips \cite{OpenPATH}. NREL also provided a Python-based vehicle energy consumption prediction engine, RouteE \cite{RouteE}. While OpenPATH and RouteE are useful for optimizing the most energy-efficient route for vehicles, they take all types of energy into account, and are not designed for the measurement of EVCL.

EVCL is produced when EVs recharging at EVCSs, and the time and place where the EVs recharge depends on their arriving time and destinations. The UE models optimizes the stable distribution of the traffic flow and EVCL under the constraints of OD pairs, unable to characterize the dynamic features under short time-scale parameters alternation \cite{dynamic} or the behavior of a single vehicle, or to say, microscopic behavior. Dynamic features are considered in \cite{apen_dynamic} while the uncertainty of traffic demand is solved by stochastic optimization. For the stochastic models (including the MC models), one of its shortcomings is that the randomness may lead to the inaccurate representations and ignore the influence of single-vehicle behavior.

A solution to the problems mentioned above may be taking the microscopic UTN (MUTN) into account. In an MUTN, a single vehicle instead of traffic flow is the fundamental unit, and the motion of each vehicle is simulated in a specific step, which may deal with the problems lying in UE models. Microscopic behavior such as car-following, lane-changing and traffic lights is also considered in an MUTN simulation, which may increase the accuracy compared to the traditional MC models. 

SUMO, a software for the microscopic simulation of UTN, was used to implement such MUTN simulation for EVCL modeling in \cite{swx_early}. It was noticed that SUMO possessed an internal EV energy consumption meter which was likely to underestimate the realistic values. Therefore, an alternative energy consumption model was employed in \cite{sumo2} to make the instant energy consumption estimation more accurate. However, these SUMO-based MUTN simulation models did not take PDN and Vehicle-to-grid (V2G) into consideration. What's more, most papers consider only one model for all EVCSs, while exisiting EVCSs can be divided into FCSs and slow charging stations (SCSs). Since the operation in FCS and SCS may be different, a single model may not describe the both charging station (CS) types accurately at the same time.

To mitigate the problems mentioned above, a platform named V2Sim is proposed in the paper. V2Sim is a Python-based simulation platform for coupled MUTN and PDN, aiming at simulating the process of EVCL generation. In this platform, SUMO is employed to simulate the microscopic behaviour in MUTN, and DistFlow is utilized to describe the optimization of the PDN. What's more, customizable V2G operation is supported, and dual CS models are designed for FCS and SCS respectively, enabling the co-simulation of both FCS and SCS. To boost the simulation, multi-processing technology is utilized to enable the parallel simulation of multiple cases, accelerating large-scale data generation. As far as we know, this is the first Python-based open-source platform, allowing V2G-compatible simulation and optimization of both MUTN and PDN. Notably, V2Sim is a highly extensible platform, supporting customization and dynamic adjustment on many modules, For example, the V2G strategies about when to enable V2G and how to allocate V2G demand to each EV can be customized, and a CS can be turn on or off dynamically during simulation. Such features greatly expand the available scenarios of the platform.

\section{Simulation Framework}
The proposed platform, V2Sim, has following featured goals:

\begin{enumerate}
    \item Simulate the EVCL generation considering both FCS and SCS in coupled MUTN and PDN.
    \item Perceive V2G influence with customizable V2G strategy.
    \item Demonstrate the dissemination and evolution of CS fault.
    \item Accelerate multiple-case simulation by multi-processing.
\end{enumerate}

\begin{figure*}[htbp]
    \centering
    \includegraphics[width=5.2in]{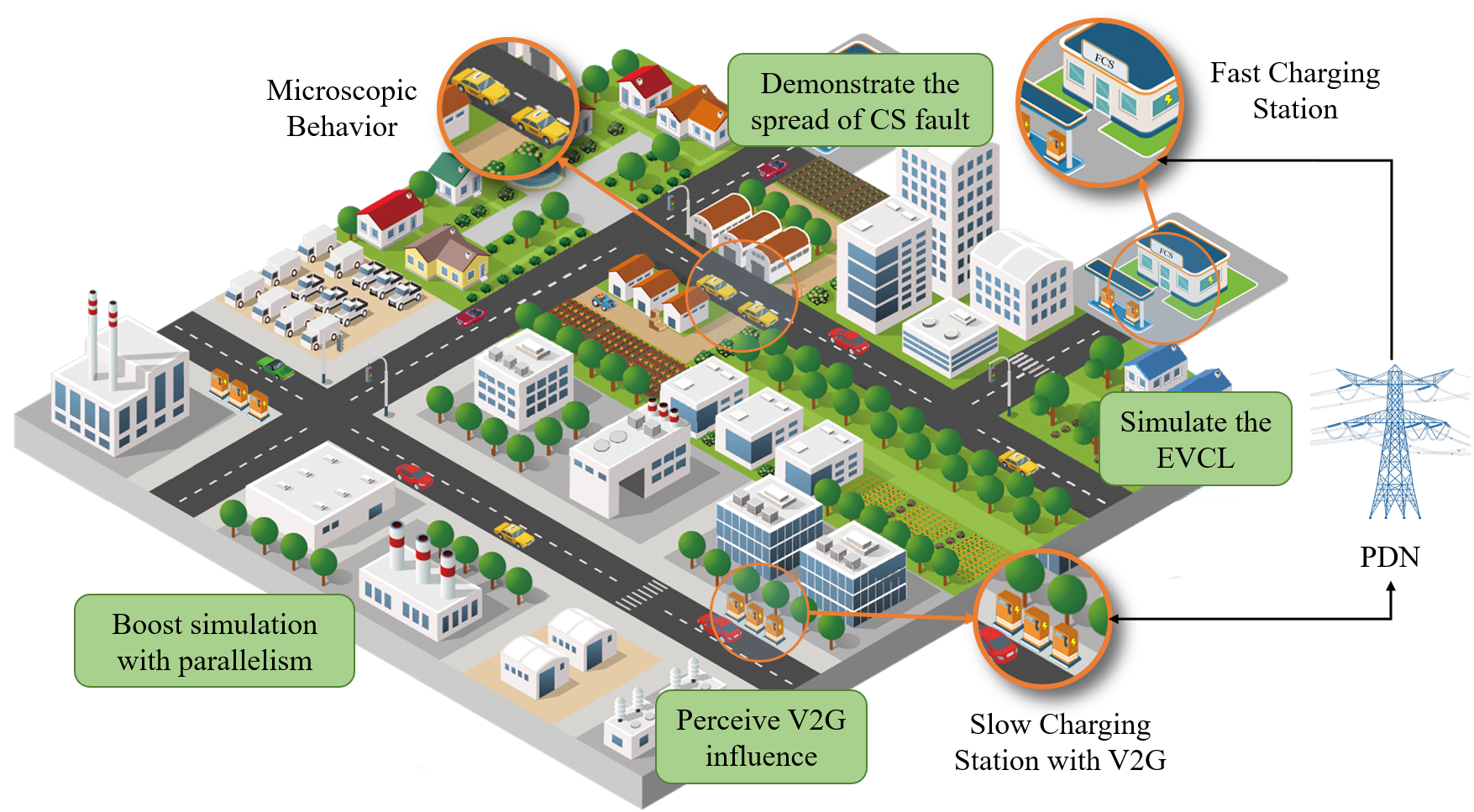}%
    \caption{Main goals of V2Sim}
    \label{fig_front}
\end{figure*}

Figure \ref{fig_front} shows the main goals of the proposed platform. Besides the major goals mentioned above, there are auxiliary functions, including the basic PDN optimization, EV properties (SoC, location, status) monitoring, and so on. A set of peripheral tools is also supplied, including EV trip generator, CS generator, plot kit, log file analyzer and result file viewer.

\begin{figure*}[htbp]
    \centering
    \subfloat[UTN Subsystem]{\includegraphics[width=5.2in]{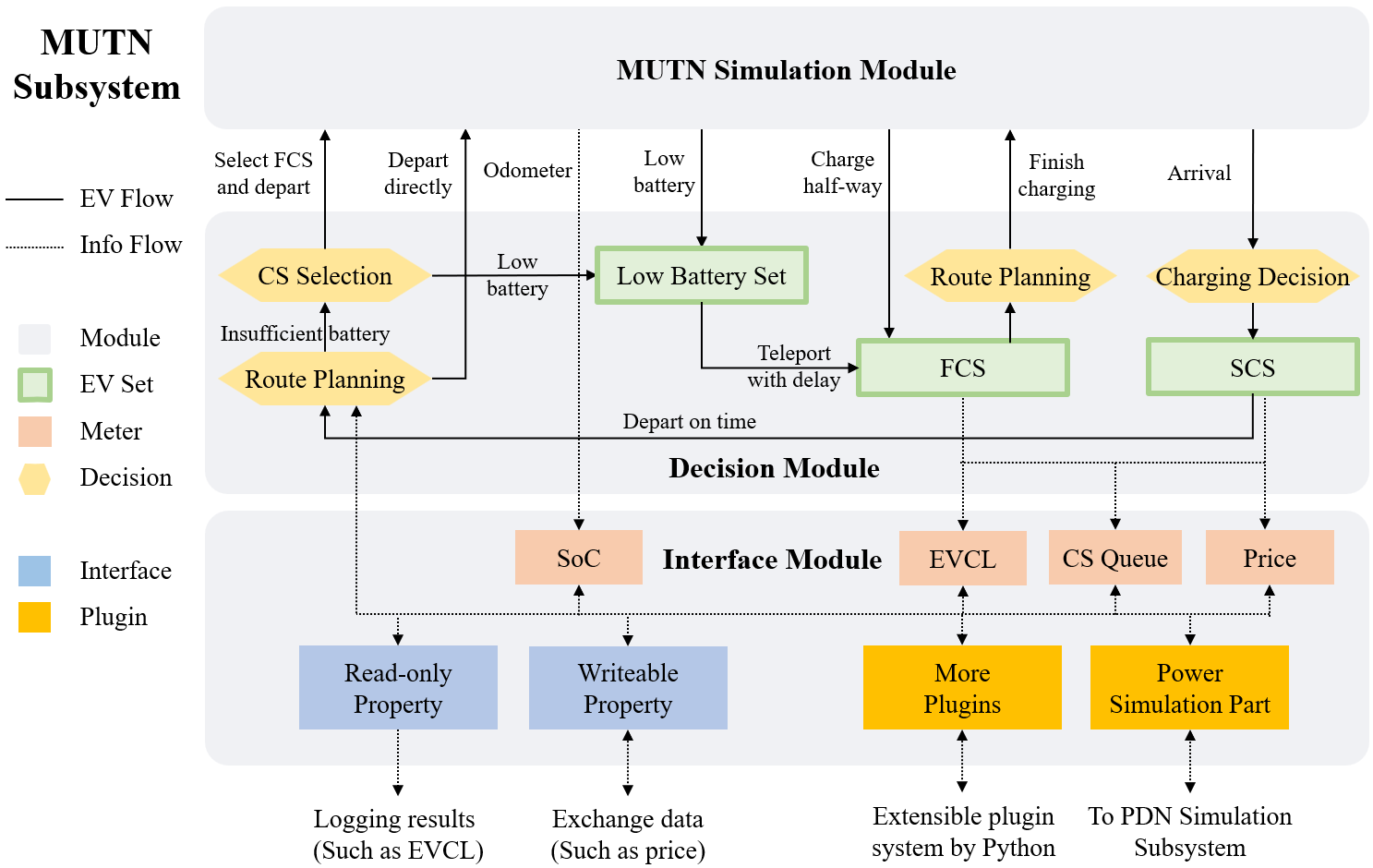}%
        \label{fig_utn}}
    \hfil
    \subfloat[PDN Subsystem]{\includegraphics[width=5.1in]{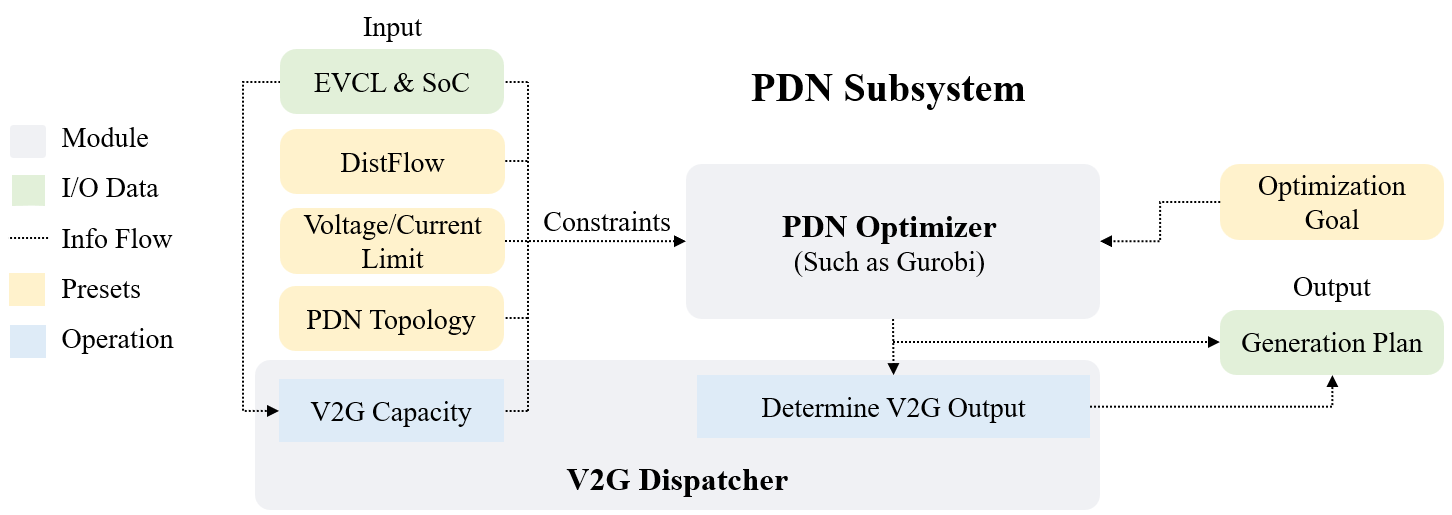}%
        \label{fig_pdn}}
    \caption{Structure of V2Sim Platform. Consists of two subsystems connected by a plugin module.}
    \label{fig_str}
\end{figure*}

Fig \ref{fig_str} illustrates how V2Sim is implemented, which comprises the MUTN and PDN subsystems. The MUTN subsystem is divided into three parts: the MUTN simulation module, the decision module, and the interface module. The PDN subsystem is separated into two parts: the V2G dispatcher and the PDN optimizer.

\subsection{MUTN Simulation Module}
This module is the core of microscopic simulation for single vehicles. In order to simulate the vehicles' motion, the MUTN and the EV are modeled. The microscopic behavior of a single EV in the MUTN is introduced. 

\subsubsection{MUTN model}
A UTN is defined as a directed graph $ G=<V, E> $, whose vertices $V$ are road junctions, and the edges $E$ are the roads. Since the graph is directed, any bi-directional road is treated as two uni-directional roads. Some bi-directional roads in the road network are dead-ended. When vehicles drive to a dead-end, they can only perform a U-turn. These dead-end points are also treated as junctions. Generally, the UTN will not change during simulation; thus, the directed graph is static. A vehicle always starts from an edge and ends at another. The optimal path algorithm is required to determine how an EV should move between an OD pair. 

\subsubsection{EV model}
The parameters of a common vehicle include acceleration capability, deceleration capability, vehicle length, and maximum vehicle speed. In this study, to generate EVCL, additional EV specific parameters were defined: battery capacity, average discharge per unit distance, charging power, charging efficiency, discharge efficiency, and V2G power.

The battery capacity characterizes the capacity of the vehicle mounted battery, measured in kWh. There is a certain relationship between battery capacity and vehicle endurance.

The average discharge per unit distance represents the amount of electricity required by an EV per unit distance traveled, measured in kWh/m. For the convenience of calculating the range, this study simplified the characteristics of the EV motor and battery, assuming that the amount of electricity per unit distance traveled by the vehicle is the same in any situation.

Charging power represents the EVCL generated by a vehicle after connecting to a charging station, in kW. Each EV has different charging powers defined for both FCS and SCS. According to the actual situation, when the SoC of EV batteries is above 70\% to 80\%, the charging power will significantly decrease. Therefore, this study adopts a segmented charging power model. When the SoC is less than or equal to 80\%, the charging power remains constant; When the SoC is greater than 80\%, the charging power linearly decays. The charging power when the SoC is less than or equal to 80\% is denoted as $P_0$, then the charging power $P$ subjects to
\begin{equation}
	p^{i}_c(x^{i})=\left\{
	\begin{array}{cccc}
		p_{c0}^{i} & 0\le x^{i}< 0.8\\
		p_{c0}^{i}(3.4-3x^{i}) & 0.8\le x^{i}\le 1
	\end{array}\right.
\end{equation}
where $x^{i}$ denotes the SoC of the EV $i$. Since this approximation may not be able to satisfy all the situations, the platform also supports customized charging power model.

V2G power describes the maximum discharging power if the EV is required to send electricity back to the power grid, also in the unit of kW. EV $i$ also has a coefficient, $k_{v}^i$, to measure the EV user's will to join V2G program. If the $SoC^{i}$ is lower than $k_{v}^i$, then the EV will not discharge V2G power. In the model employed, an EV shall never generate V2G power autonomously. The necessary conditions for an EV to generate V2G power are:
\begin{itemize}
	\item{$SoC^{i} \ge k_{v}^{i}$;}
	\item{Receive the central dispatcher's command.}
\end{itemize}

\subsubsection{Microscopic behavior}
SUMO employs microscopic models to simulate the movement of every single vehicle on the street, mostly assuming that the behavior of the vehicle depends on the vehicle's physical abilities to move and the driver's controlling behavior\cite{sumo}. The car-following and lane-change models are two important SUMO components used in this module.

The car-following mode\cite{car_follow} is constructed under the assumption of no crash. Each EV has a ideal maximum speed $v_{max}$ and a safe speed avoiding crash $v_{safe}$. The acceleration of the EV lies in a given range $[-b,a](a,b>0)$. The speed of an EV subjects to 
\begin{equation}
	v(t+\Delta t)=\max\left\{0,\min\left\{v_{max},v(t)+a\Delta t, v_{safe}\right\}-\eta\right\}
\end{equation}
\begin{equation}
	v_{safe}=v_l(t)+\frac{g(t)-v_l(t)t_r}{\frac{v_l(t)+v_f(t)}{2b}+t_r}
\end{equation}
where $v_l(t)$ denotes the speed of the former vehicle at time $t$, $g(t)$ denotes the distance between current EV and the former EV, $t_r$ denotes the reaction time of the driver, $v_f(t)$ denotes current vehicle speed, $b$ denotes the maximum deceleration of the vehicle and $\eta$ describes the random speed decrease caused by the not idealized driver operation.

The lane-change model is proposed in \cite{lane_change}. The basic idea of the lane-change model is that the vehicle only changes lanes when there is sufficient physical space in the target lane. The lane-change model consists of four steps: 
\begin{enumerate}
	\item{Determine the optimal lane that can be changed;}
	\item{Calculate the safe vehicle speed for the current lane based on the speed requirements related to the previous simulation and lane changing;}
	\item{Evaluate lane changing requirements;}
	\item{Execute the lane change action or calculate the speed of the next simulation step according to the urgency of the lane change request.}
\end{enumerate}

\subsection{Decision Module}
This module make critical decision for departure and CS selection. It also maintains the procedure of EV recharging and counts the EVs depleting battery.

\subsubsection{CS model and low battery set}
Two types of CS, the slow CS (SCS) and the FCS, are attached to the road network. An SCS is attached to each uni-directional road. Two SCSs are attached to a bi-directional road, since it is viewed as two uni-directional roads. A bi-directional dead-end road can be marked as an FCS, and no SCS will be attached to this road. From the description above, an EV trip can be seen as a trip from one SCS to another, possibly stopping temporarily at no more than one FCS. 

Whatever the type of CS, it is always equipped with a given but runtime-changeable number of charging piles. The money an EV user purchases when buying a unit of electricity from the CS is defined as the {\it user purchase price} (UPP) of this CS. For various reasons, a CS may not be able to serve users all the time. When a CS can serve the users, it is {\it online}; otherwise, it is {\it offline}. UPP and online status of a CS can be defined in the configuration file or set dynamically at runtime.

The FCS and the SCS follow different charging patterns. In an FCS, once an EV finishes charging, it will leave immediately and free the pile it occupied. If all the piles are occupied, then extra vehicles will queue and follow the principle of `first come, first served'. In an SCS, an EV occupying a pile shall never leave until all the piles are occupied; then, no more vehicles will be able to get charged here, even if they are willing to queue.

If an EV depletes its battery during driving, it will be removed from the simulation immediately and sent to the low battery set. After twice the time it needs normally driving to the nearest FCS, it will be teleported to the that FCS.

\subsubsection{Route Planning}
There are three algorithms for finding the optimal path inside SUMO: Dijkstra, A*, and Contract Hierarchy (CH) \cite{sumo}. They produce the same results with different efficiency. There will be little difference among the three algorithms in a small road network with only several dozens of junctions. When a real-world road network is adopted, the CH algorithm will perform best, A* the second and Dijkstra the worst. Notably, the CH algorithm needs pre-processing, which may take some time. So, if only the optimal path is calculated just a few times, Dijkstra and A* will be better, even if in an extensive road network.

When an EV departs from one edge, its route should be determined. The route selected for the EV is the optimal path between the EV's OD pair. Here, `optimal' can be either shortest or fastest. The shortest paths are solved when the edge weight is the road length. Since the road length is unchanged, the shortest path between any OD pair needs to be solved no more than once. The fastest paths at a specific time are solved when the edge weight is the road's average passing time (APT). As the APT of a road is dynamic, the fastest path of a particular OD pair must be re-calculated if it is used at different times.

When an EV departs from an SCS, there are two strategies available for the departure decision, namely `threshold strategy' and `distance strategy'.

Under the threshold strategy, a threshold for SoC, $k_f^i$, is defined for each EV $i$ respectively. If the SoC is less than $k_f^i$, EV $i$ will first visit an FCS to get fully charged and then go to the destination. The selection of FCS will be introduced in the next section.

Before introduce the distance strategy, we define that a destination is {\bf reachable} for EV $i$, if its user believes EV $i$ is able to arrive the CS. Formally, a place is reachable for EV$i$ if $k_{r}^iL\le d^i$, where $k_{r}^i$ is a coefficient depicting the user preference, $L$ denotes the length of the fastest path between the OD pair, and $d^i$ denotes the range of EV $i$ with electricity remaining. Otherwise, if $k_{r}^iL > d^i$, then the place is {\bf unreachable} for EV $i$.

Under the distance strategy, the length of the fastest path between the OD pair is measured. If the destination is unreachable for an EV, it will firstly visit an FCS to get fully charged, and then go for the destination. Otherwise, it will directly go to the destination.

Under both strategies, if the EV sets out without a halfway FCS given, it will go directly to the destination along the fastest path determined at the time of its departure. The route of the EV shall never change half-way. The strategy for departure decision can be altered runtime by a global switch.

When an EV finishes charging and departs from an FCS, its route will be set as the fastest path from the FCS to its destination.

\subsubsection{FCS Selection}
We define a CS is {\bf nearby} for an EV $i$ if the Euclid distance of the EV and the CS is smaller than a given threshold. In the case of FCS selection, the scores of nearby online and reachable CS are calculated. For a CS $j$ and a specific EV $i$, the score $f(j)$ is defined as 

\begin{equation}
	f(j) = \omega^i(T_d^i(j)+n_w(j)T_w)+c^i\Delta W^i
\end{equation}
where $\omega^i$ is the coefficient of $EV$ i to describe the value of unit time of its user. $T_d^i(j)$ denotes the time needed to travel from the current position of EV $i$ to CS $j$. $n_w(j)$ denotes the number of waiting vehicles in CS $j$ and $T_w$ is a given constant describing the average time of fully charging an EV. $c^i$ represents the unit cost of electricity and $W^i$ denotes how much electricity will be charged into EV $i$.

The CS to be selected will be the CS with the minimum score. If no CS is reachable and online, the EV will be removed from the simulation and marked `low battery'.

\subsubsection{Charging decision}
A threshold for SoC, $k_s$, is defined for each EV respectively. On the arrival of an EV, the module will check whether its SoC is greater than or equal to $k_s$. If the SoC is less than $k_s$, the EV will try to join the SCS at the destination on condition that at least one free pile exists; otherwise, the EV will park at the destination without occupying any pile in the SCS.

\subsection{Interface Module}
This module serves as the interface for reading the simulation parameters and modifying some parameters externally, which enables data logging. 

This module also create a plugin system with extensibility. Any step-based function can be programmed by Python added as external plugins to this platform, making user-defined custom functions available to this platform. For example, the PDN module of the platform is implemented as two plugins, the PDN simulation plugin and the V2G plugin. Each plugin has three phase, the initialization phase, the pre-step phase and the post-step phase. The initialization phase will be executed only once before the simulation start. The pre-step phase will be executed before each simulation step and the post-step will be executed after each simulation step. The execution sequence of the plugins adheres to the sequence they are added. Dependency relations among the plugins are allowed in the plugin system. For example, the V2G plugin cannot work without the PDN simulation plugin.

\subsection{The PDN Optimizer}
The PDN optimizer aims at optimizing a given objective and produce corresponding generation plan, under the constraints of volatge and current. DistFlow model was used to solve the optimization problem. Formally,

\begin{equation}
	\min \sum_{i}c^iP_i
\end{equation}
s.t.
\begin{equation}
	\begin{split}
		\sum\limits_{i:i \to j} {({P_{ij}} - {R_{ij}}{l_{ij}})} + P_j^G &= \sum\limits_{k:j \to k} {{P_{jk}}}  + P_j^D,j \in B \\
		\sum\limits_{i:i \to j} {({Q_{ij}} - {X_{ij}}{l_{ij}})} + Q_j^G &= \sum\limits_{k:j \to k} {{Q_{jk}}}  + Q_j^D,j \in B\\
		{v_j} = {v_i} - 2({R_{ij}}{P_{ij}} + {X_{ij}}{Q_{ij}}) &+ (R_{ij}^2 + X_{ij}^2){l_{ij}}, \left<i,j\right>\in L\\
		P_{ij}^2 + Q_{ij}^2 &\le {l_{ij}}{v_i},\left<i,j\right>\in L
	\end{split}
\end{equation}
where $B$ denotes the bus set, $L$ denotes the transmission line set, $\left<i,j\right>$ denotes a line starts from bus $i$ and end at bus $j$, $P_ij$ and $Q_ij$ denote the active and reactive power transmitted from bus $i$ to bus $j$, $P_j^D$ and $Q_j^D$ denote the active and reactive power load at bus $j$, and $P_j^G$ and $Q_j^G$ denote the active and reactive power generation at bus $j$. These constraints is determined by the PDN topology, EVCL. Besides, the voltage of each bus and the current of each line is also limited.

The UTN simulation and the PDN simulation can be asynchronous and may be assigned different time steps. When the PDN load does not change drastically, it will be acceptable to simulate PDN with a larger time step than the UTN simulation.

\subsection{The V2G Dispatcher}
In the proposed platform, the minimum V2G dispatching unit is SCS, and the V2G output is only enabled at a given period of time instead of all day long. Each SCS, if enabled V2G, is viewed as both a load and a generator in the PDN, and the generation plan produced in the PDN optimizer also includes these SCSs.

The V2G dispatcher collects the maximum V2G power each SCS may supply and allocate the planned power to supply to each SCS. For SCS $j\in C_s$, the V2G capacity $P_{vc}^j$ is defined by 
\begin{equation}
	P_{vc}^{j}=\sum_{i\in j} [SoC^{i}\ge k^i_{v}]p^{i}_{v}
	\label{form_v2g}
\end{equation}
where $[x]$ is a binary condition, $[x]=1$ if condition $x$ is true, $[x]=0$ if condition $x$ is false. 

If an SCS is assigned a given number of V2G output power, it will be distributed equally to each EV willing to participate in V2G program by the default strategy.

When the V2G dispatcher works, each CS will submit its capacity and wait for a response from the dispatcher. The default dispatching strategy takes electricity evenly from EVs that willing to participate in V2G. Formally, denote the actual V2G power required in the response as $P_{vcr}^j$, then each EV willing to participate in V2G will output a V2G power of $p^{i}_{v}P_{vcr}^j/P_{vc}^j$ by default. It is assured that $0\le P_{vcr}^j\le P_{vc}^j$. Customized V2G strategies are allowed to make the platform more usable.

\section{Platform Usage}
To utilize V2Sim, standard steps should be followed. Figure \ref{fig_flow} shows the standard steps, which is explained below:

\begin{figure}[htbp]
\centering
\includegraphics[width=3.5in]{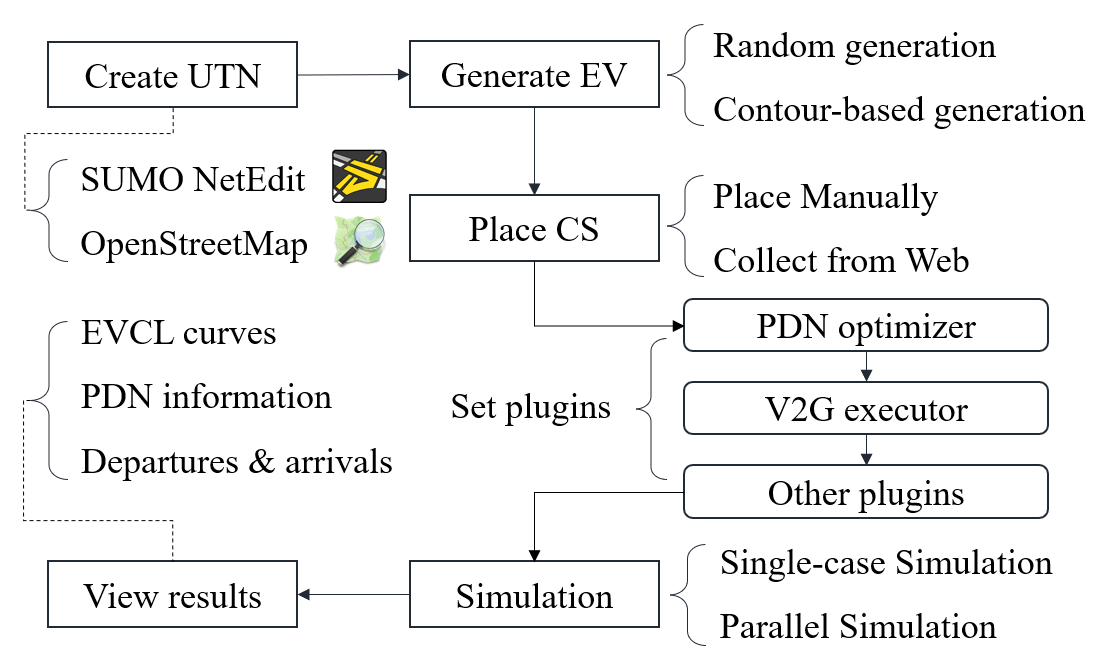}\caption{Standard steps.}
\label{fig_flow}
\end{figure}

\subsection{Create UTN}
To simulate the behavior of EVs, a road network must be created. There are two recommended ways to create a road network. The first option is to create a simplified network from scratch in SUMO NetEdit, which is convenient for simple virtual road network. Another recommended option is to employ the road network create tool provided by SUMO. By selecting the desired area on OpenStreetMap, a real-world road network is downloaded and created.

\subsection{Generate EVs, trips and CSs}
After the road network is created, EVs and their trips should be generated to specify their parameters, such as the $k_f^i,k_s^i,k_r^i$ of EV $i$.

Trips, or traffic demand $D$, can be described as a set, whose elements are OD pairs. There may be multiple trips for a single EV. A trip $\theta$ can be denoted by a triplet $(T,u,v)$, denoting a trip from $u_1$ to $u_2$ departing at time $T$. Suppose there are $n$ trips for a single EV, and the destination of the last trip must be exactly the origin of the first trip, then the $n$ trips can be denoted as 
\begin{equation}
	\left\{
	\begin{array}{ll}
		\theta_1&=(T_1,u_1,u_2) \\
		\theta_2&=(T_2,u_2,u_3) \\
		... & \\
		\theta_{n-1}&=(T_{n-1},u_{n-1},u_n)\\
		\theta_n&=(T_n,u_n,u_1)
	\end{array}\right.
\end{equation}
where $T_1<T_2<...<T_{n-1}<T_n, T_1-c\sim\Gamma(a,b)$, $a,b,c$ are all given constants. $\forall i\in \mathbf N_+,i<n,T_{i+1}-T_i$ assumes a specific distribution. The transition of trips $\theta_1,\theta_2,...,\theta_{n-1}$ by $u_1,u_2,...,u_n$ can be viewed as Markov processes.

\subsection{Place CS}
There are two methods to place CS in the road network. In the first method, edges whose ID start with ``CS'' are recognized FCSs while other are SCSs. In the second method, FCSs are determined by the POI positions collected from Amap, while SCSs are placed according to the building contour collected from OpenStreetMap. Other parameters, like the number of charging piles in a CS, should also be specified.

\subsection{Set plugins}
To optimize PDN and perceive V2G, corresponding plugins must be enabled. Besides, any other customized function should be coded as plugins to insert into the simulation. For the user-defined EV charging feature and V2G strategy, they should be defined in the additional code, a Python file working together with the simulation configuration. The PDN optimizer is only used for how much V2G power should be supplied by the SCS. It can be replaced by any other plugins which have the identical function.

\subsection{Simulation}
Simulation gets started when the steps above are done. Single case simulation or parallel simulation for multiple cases are available choices. By configuring the command line parameters, various parameters can be altered to conduct comparative experiments.

\subsection{View results}
By using the plot kits provided, the results can be drawn in figures to demonstrate the fluctuation and trend of the data. Drawing presets are provided to plot frequently-used figures, while advanced plot terminal is also supplied for drawing customized figures.

\section{Case Analysis}
In this section, a 37-junction simplified road network and a real-world road network are analyzed to verify the ability of the proposed platform to simulate the UTN and PDN. Then, the acceleration of this proposed platform is briefly introduced. Finally, a UE model is taken for comparison with the proposed platform.

\subsection{37-junction simplified case}
\subsubsection{Case setup}
\begin{figure}[htbp]
\centering
\includegraphics[width=2.8in]{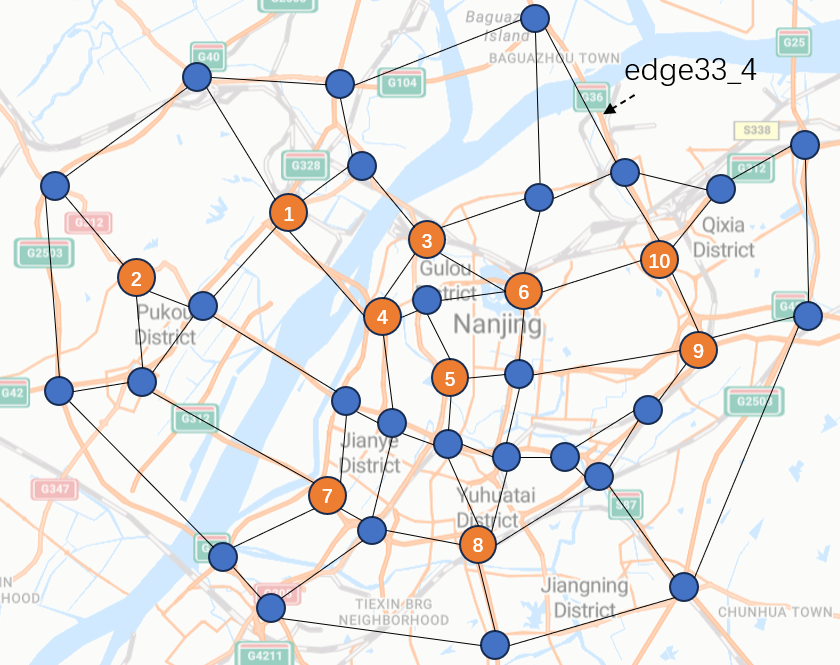}\caption{37-junction road network.}
\label{fig_nj}
\end{figure}

The topology of 37-junction road network is shown in Fig \ref{fig_nj}. This network, containing 37 junctions and 65 bi-directional roads, is a simplification of a real-world UTN located in Nanjing, China. 10 FCSs, whose numbers are identify in Fig \ref{fig_nj}, are placed in the road network. Besides, 10 charging piles are placed in each CS, whatever FCS or SCS.

The IEEE 33-bus PDN\cite{ieee33} is adopted in this case, which contain 33 buses and 32 transmission lines. Four generators are placed at bus 2, 3, 6, 8 respectively. Bus 1 is connected to the external power grid, which is also treated as a generator. All the generators' cost follow $f(x)=0.0001x^2+0.3x+10$ (unit of $x$ is kWh, and unit of $f(x)$ is dollar), while the money paid to user for purchasing V2G power is \$1.0 per kWh.

5000 EVs are added into the road network for simulation. All the EVs are sampled from 6 prototypes, listed in Table \ref{tab_pro}. The initial SoC assumes certain distribution, with an average about 0.60. The property $\omega^i,k^i_r,k_s^i,k_f^i,k^i_v$ of EV $i$ all assume certain distributions, specifically $\omega^i\sim U(5,10),k^i_r\sim U(1.0,1.2),k_s^i\sim U(0.4,0.6),k_f^i\sim U(0.2,0.25),k^i_v\sim U(0.65,0.75)$, where $U$ refers to uniform distribution.

\begin{table}[htbp]
\begin{center}
\caption{EV prototype}
\label{tab_pro}
\begin{tabular}{ccccc}
\hline
ID & \makecell[cc]{Battery \\capacity/kWh} & \makecell[cc]{Discharge \\rate/Wh/m} & \makecell[cc]{Fast charging \\power/kW} & \makecell[cc]{Slow charging \\power/kW} \\
\hline
P1 & 100 & 0.159 & 200 & 5.98 \\
P2 & 55.9 & 0.151 & 60 & 7 \\
P3 & 84 & 0.210 & 7 & 7 \\
P4 & 76.8 & 0.171 & 100 & 7 \\
P5 & 90.3 & 0.181 & 60 & 7 \\
P6 & 100 & 0.196 & 100 & 7 \\
\hline 
\end{tabular}
\end{center}
\end{table}

In the case proposed, the simulation will last for 7 days. In the initial state, the road network is empty and the charging status is randomly generated instead of employing a real situation; therefore, an extra day of simulation is added before the formal simulation to make the state of EVs and road network more real. For each EV, 3 trips are generated for each day. Since the simulation will continue for 8 days, therefore, 24 trips are generated for each EV. On weekdays, the first trip departure time $T_1$ assumes: $T_1-114.54\sim\Gamma(6.63,65.76)$; on weekends, the first trip departure time $T_1$ assumes: $T_1-197.53\sim\Gamma(3.45,84.37)$. The interval between trips assumes certain distribution, determined by the type of origins. The probability of Markov transition among the places is also determined by the type of origins. The detailed data are shown in the appendix.

The following sections will demonstrate the results.

\subsubsection{EVCL considering both FCS and SCS}
\begin{figure}[htbp]
	\centering
	\subfloat[EVCL at FCSs]{\includegraphics[width=3.5in]{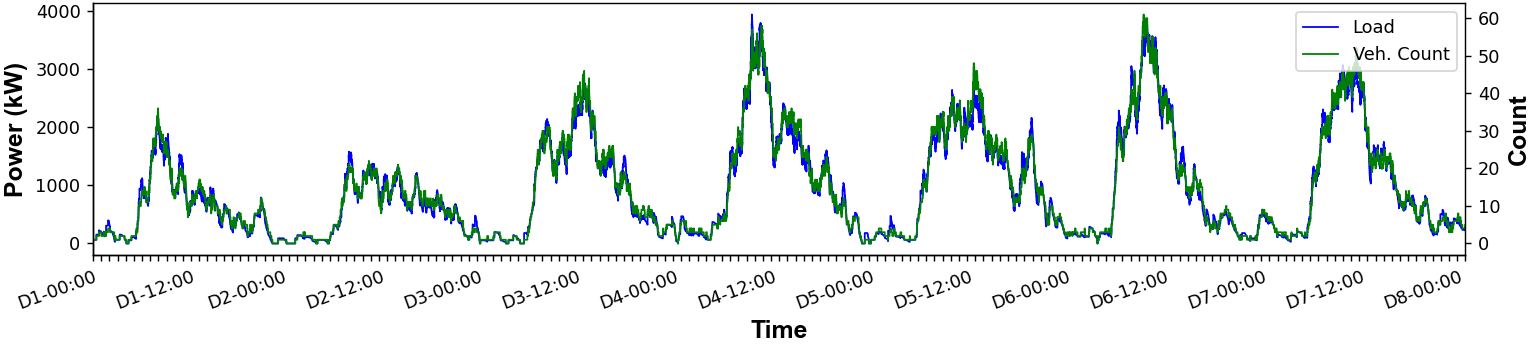}%
		\label{fig_whole_fcs}}
	\hfil
	\subfloat[EVCL at SCSs]{\includegraphics[width=3.5in]{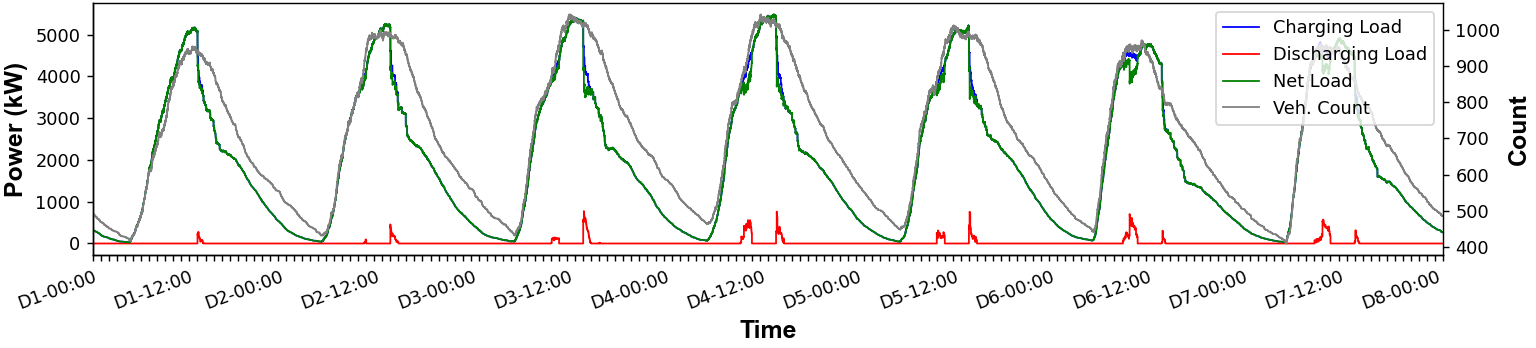}%
		\label{fig_whole_scs}}
	\caption{Simulation results of a whole week with V2G.}
	\label{whole_week}
\end{figure}

Figure \ref{whole_week} shows the simulation results of EVCL at FCS and SCS during a whole week. As the parameters for weekday and weekend are different, it can be found that the FCS PL at weekend is higher than that at weekdays, while the SCS PL at weekend is lower than that at weekdays. Data shown in Figure \ref{whole_week} does not enable V2G function.

\subsubsection{V2G awareness}
\begin{figure}[htbp]
    \centering
    \subfloat[With V2G]{\includegraphics[width=3.5in]{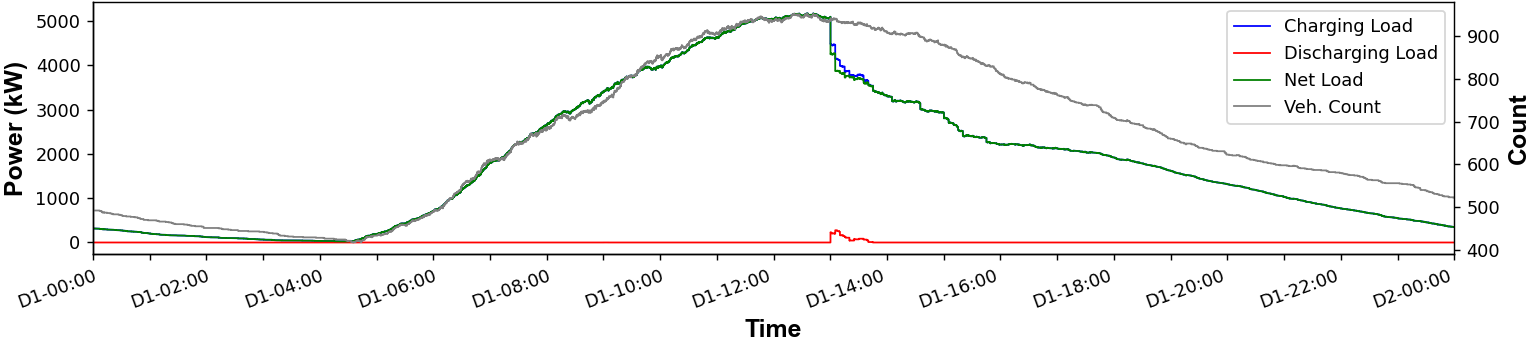}%
        \label{fig_cmp_slow_v2g}}
    \hfil
    \subfloat[Without V2G]{\includegraphics[width=3.5in]{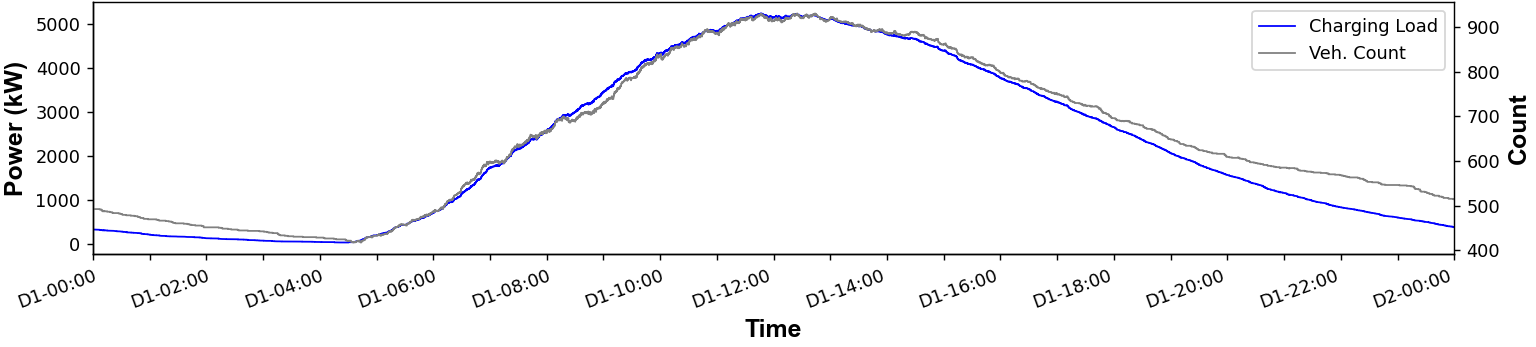}%
        \label{fig_cmp_slow_nov2g}}
    \caption{Simulation results of SCS load with and without V2G}
    \label{fig_vvv}
\end{figure}

Figure \ref{fig_vvv} displays the case results with and without V2G respectively. In this case, V2G is enabled at 8:00-10:00 and 13:00-16:00. It can be discovered that when V2G is enabled, a plunge appears in the accumulated slow charging load.

The first plunge happens around a quarter to ten when the cost of unit V2G output becomes lower than the cost of the unit output of a normal generator. Therefore, to minimize the cost (including the cost of V2G output and normal generator output), V2G power begins to be utilized, causing a sudden drop in slow charging load.

The second plunge happens at the beginning of the V2G period in the afternoon. At this time, the cost of unit V2G output is still lower than the cost of unit output of a normal generator. Therefore, the V2G power is used immediately. However, since the V2G power is quite large for each EV (about 20kW per EV), the SoC of EV declines quickly, and soon it will be lower than the threshold $k_v^i$. Thus, the V2G capacity is drained quickly. It is notable that the slow charging load does not resume to the level before V2G is enabled due to the charging limits of V2G. When V2G is disabled, an EV tries to get fully charged at an SCS; when V2G is enabled, an EV $i$ only tries to get charged when $SoC^i<k_v^i$. 

\begin{figure}[htbp]
	\centering
	\subfloat[With V2G]{\includegraphics[width=3.5in]{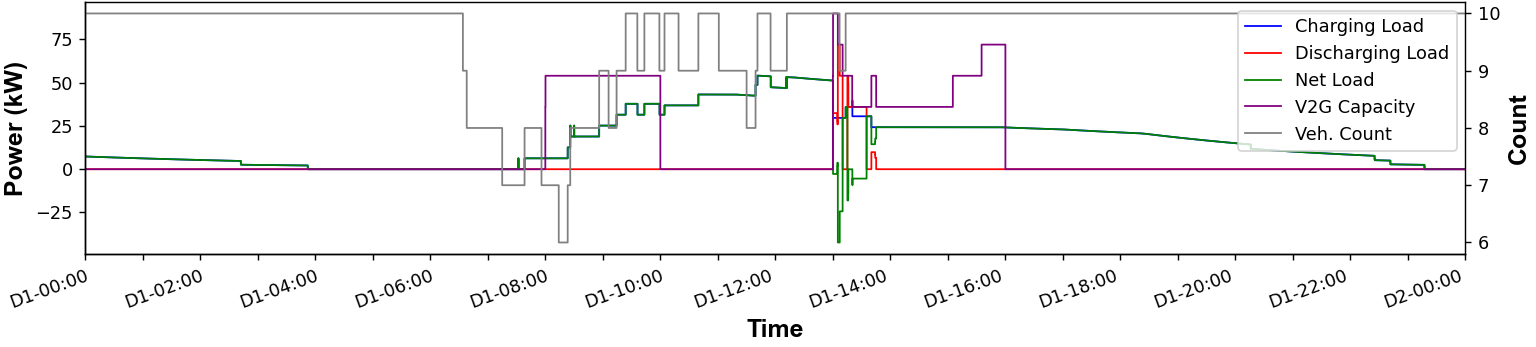}%
		\label{fig_edge_v2g}}
	\hfil
	\subfloat[Without V2G]{\includegraphics[width=3.5in]{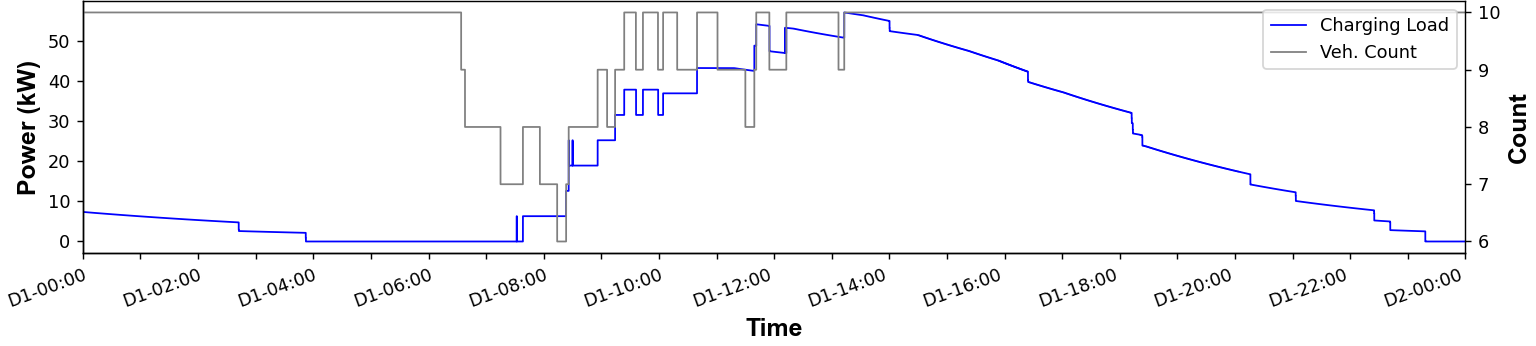}%
		\label{fig_edge_nov2g}}
	\caption{Parameters of CS egde33\_4 with and without V2G.}%
	\label{fig_vvv1}
\end{figure}

Figure \ref{fig_edge_v2g} and \ref{fig_edge_nov2g} display the parameters of SCS edge33\_4. It can be discovered that the capacity of V2G is sufficient all the time, but not utilized effectively even in the V2G enabled period. This may be a result of the comparatively high cost of V2G power. If the cost of V2G power is set lower, the utilization rate of V2G capacity may increase.

\subsubsection{Perception of CS fault dissemination}
\begin{figure}[htbp]
    \centering
    \includegraphics[width=3.5in]{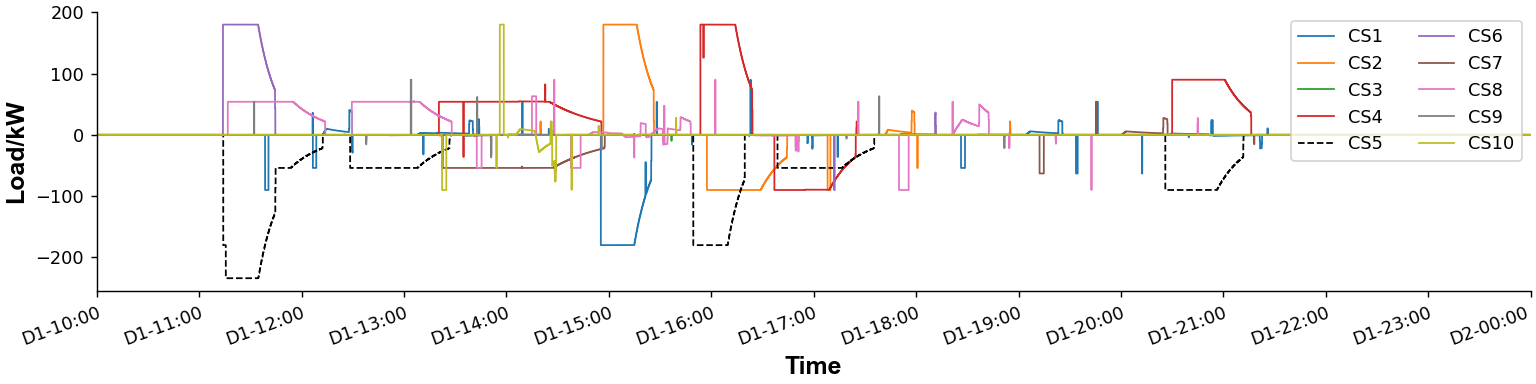}%
    \caption{Load difference when CS5 is offline since 11 a.m.}%
    \label{fig_fault}
\end{figure}

To illustrate how the platform perceives CS fault dissemination, CS5 is forced offline at 11 am on the first day to represent a CS fault. The difference of FCS load are shown in Figure \ref{fig_fault}.

To discriminate the faulted CS5, its curve is drawn in dotted lines. It can be found that the load CS5 should have taken are transferred to nearby CSs, mainly CS8, CS6, CS4, and CS2. In contrast, the remote CSs are less affected. A interesting phenomenon is that the load of certain CSs declines for a comparatively long period, like CS1, CS2, and CS7. This may be attributed to the position change of the vehicles, which leads to different CS selection.
Temporal alternation may also advance or postpone the time of load appearance, and therefore lead to a difference.

\subsubsection{Influence of CS charging price}
\begin{figure}[htbp]
	\centering
	\subfloat[Both \$1.5/kWh]{\includegraphics[width=3.5in]{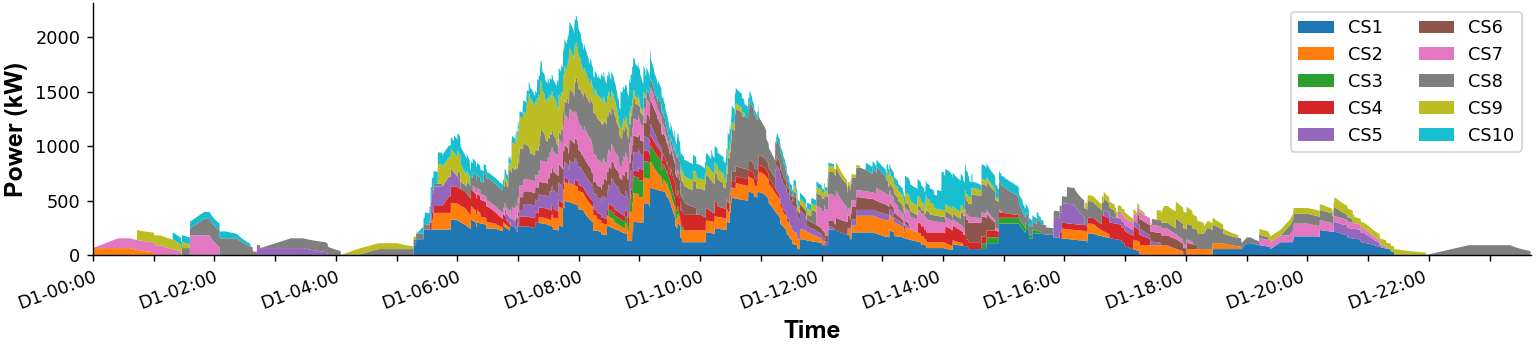}%
		\label{fig_price1}}
	\hfil
	\subfloat[Group A \$1.0/kWh, Group B \$1.5/kWh]{\includegraphics[width=3.5in]{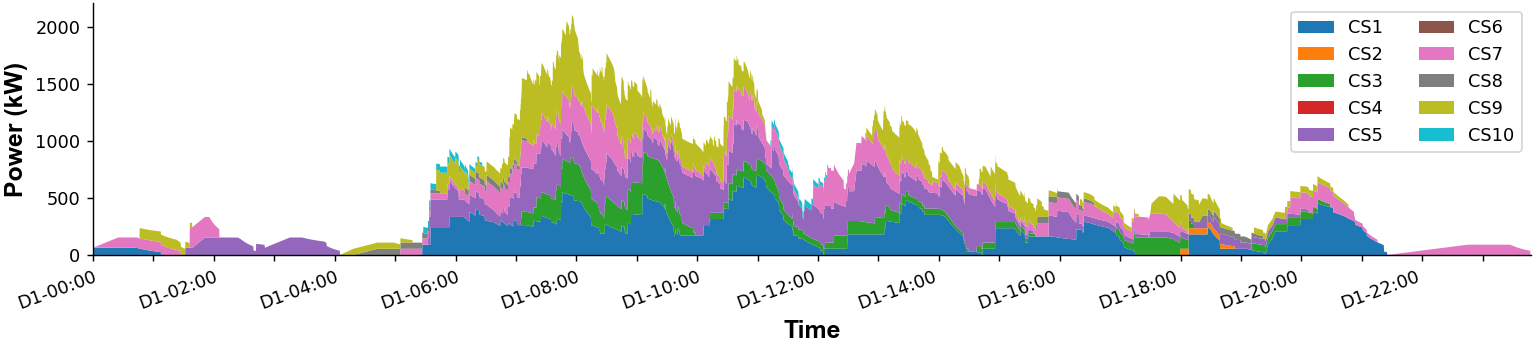}%
		\label{fig_price2}}
	\caption{Simulation results with different FCS charging price}%
	\label{fig_price}
\end{figure}

Since the selection of SCS is only related to the destination, only the influence of FCS charging price variation is considered in this section. Figure \ref{fig_price} shows the simulation results with different FCS charging price. FCS are divided into  two groups, with Group A including CS1, CS3, CS5, CS7, and CS 9, Group B including CS2, CS4, CS6, CS8, CS10. Figure \ref{fig_price1} displays the result when the FCS charging prices of both Group A and B are \$1.5/kWh, while Figure \ref{fig_price2} displays the result when charging price of Group A is \$1.0/kWh and charging price of Group B is \$1.5/kWh.

Average EVCL is employed for evaluation between Group A and B. In Figure \ref{fig_price1}, the average EVCL of Group A is 274.7kW, while the average EVCL of Group B is 277.2kW. which displays no significant difference. In Figure \ref{fig_price2}, the average EVCL of Group B is 13.8kW. Meanwhile, the average EVCL of Group A is 563.5kW, about 41 times of the average EVCL of Group B. It proves that the FCS charging price has a significant influence on the distribution of EVCL. Lower charging price will guide the users to gather in certain FCSs.

\subsection{Real-world Nanjing case}
\begin{figure}[htbp]
	\centering
	\includegraphics[width=3in]{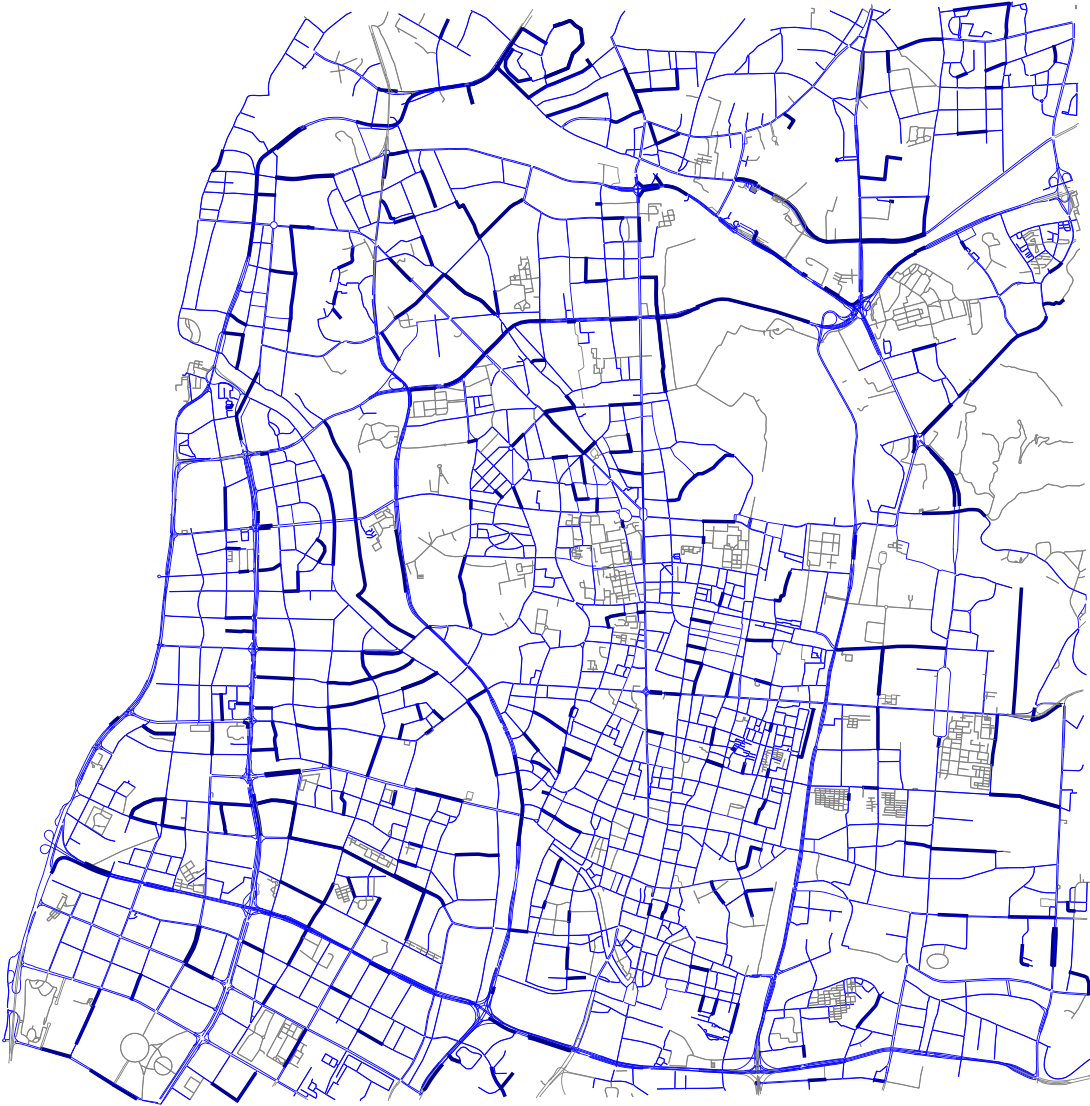}\caption{Real-world road network of Nanjing (Partial). Blue lines refer to roads. Gray lines refer to unreachable roads or non-vehicle roads. Dark blue bold lines refer toroads with FCSs.}
	
	\label{fig_real_NJ}
\end{figure}

In this section, a real-world case based on Nanjing is constructed. A partial road network of Nanjing is downloaded from OpenStreetMap, including the road topology and the building contours. 336 FCSs are added by the position searched on AMap, and 3083 SCSs are added by the building contours. 10 charging piles are placed in each CS. Figure \ref{fig_real_NJ} demonstrates the road network.

Still, the IEEE 33-bus PDN is employed, but copied for 30 times to accommodate heavier EVCL, since 100,000 EVs are added in the road network for simulation. Any parameters not referred are same as the parameter of the 37-junction case.
The simulation result of the EVCL is shown below in Figure \ref{fig_real_res}. Considering the large scale of this case, the simulation only proceeded for 2 days. The first day is used for approximating real situation while the second day for evaluating the statistics.

\begin{figure}[htbp]
	\centering
	\subfloat[EVCL of FCSs]{\includegraphics[width=3.5in]{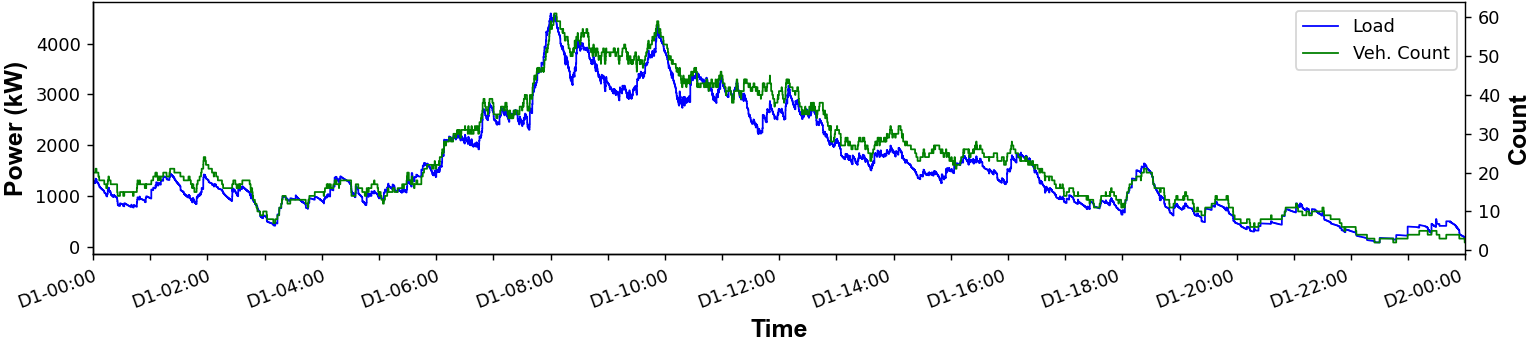}%
		\label{fig_real_fcs}}
	\hfil
	\subfloat[EVCL of SCSs]{\includegraphics[width=3.5in]{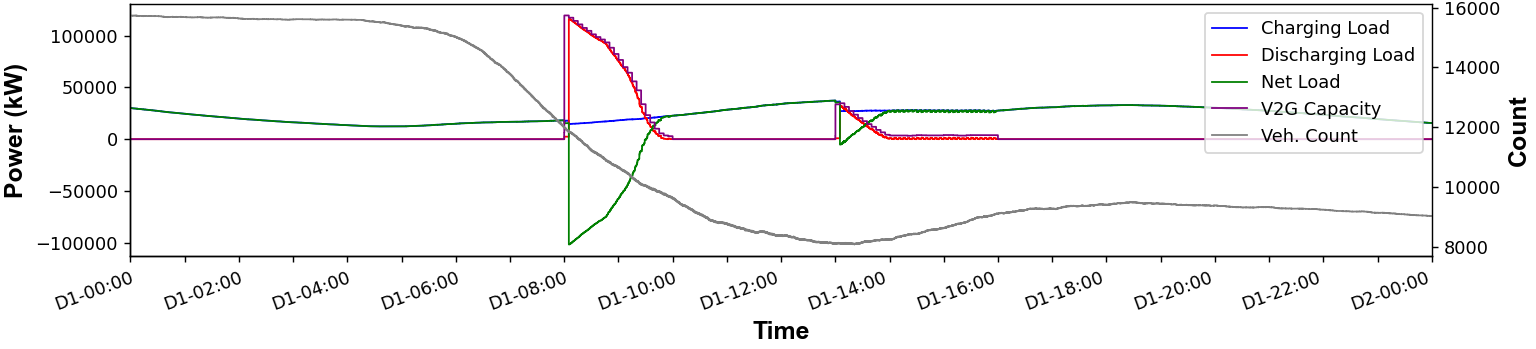}%
		\label{fig_real_scs}}
	\caption{Simulation results of the real-world Nanjing case}%
	\label{fig_real_res}
\end{figure}

\subsection{Acceleration by parallel simulation}
Since the MUTN simulation in SUMO is single-threaded, the serial simulation of multiple cases are quite slow. Thus, V2Sim makes some progress. It supports parallel simulation for multiple cases by using multi-processing technology. Experiments are conducted to measure the advantage of parallel simulation by comparing the time used of simulating the 37-junction case for 2 simulation days. The result is shown below:
\begin{enumerate}
    \item One case is simulated on a 16-core laptop, consuming 196 seconds.
    \item 16 cases is parallel-simulated on the same laptop, consuming 734 seconds.
\end{enumerate}
The 16 cases used in the second step are exactly the same as the case used in the first step, expect the random seed. It can be found that the 16 times of the simulation time of first step is far more than the second step, about $196\times 16 / 734=4.27$ times. The parallel simulation reduces the time consumed by 76.6\%. It is notable that the simulation time of the two steps are not exactly the same, which may be a result of the following factors:
\begin{enumerate}
	\item Data are recorded on the disk while simulation. The disk operation are not parallel and needs queuing.
	\item The PDN optimizer is originally a multi-threaded program which is able to utilize multiple cores efficiently. Using multi-processing technology will not bring extra advantages.
\end{enumerate}
This parallel simulation tool helps save time when performing multiple cases at the same time. It may be helpful in the following scenarios: 
\begin{enumerate}
	\item Conduct repeated experiments (only different in the random seeds) to get average metrics.
	\item Conduct contrast experiments of different cases, whose difference is the alternation of specific parameters.
\end{enumerate}

\subsection{Comparison with UE models}
A comparison with UE models is conducted to evaluate the efficiency of the simulation method. The TAP-UE method in \cite{QT_UE} is used to compare with V2Sim. In \cite{QT_UE}, the TAP-UE method produces the flow of links and EVCSs, and the EVCL at each EVCS. Since there are only FCSs consider in TAP-UE, and SCSs are ignored, therefore, SCSs are not consider in V2Sim in this comparison, either.

When 5000 EVs are added into the 37-junction road network, during a period of 48 hours, the TAP-UE gives out a solution every hour (i.e. 48 solutions for the 48 hours), and V2Sim produces continuous results.

\begin{figure}[htbp]
    \centering
    \subfloat[Proposed platform]{\includegraphics[width=3.5in]{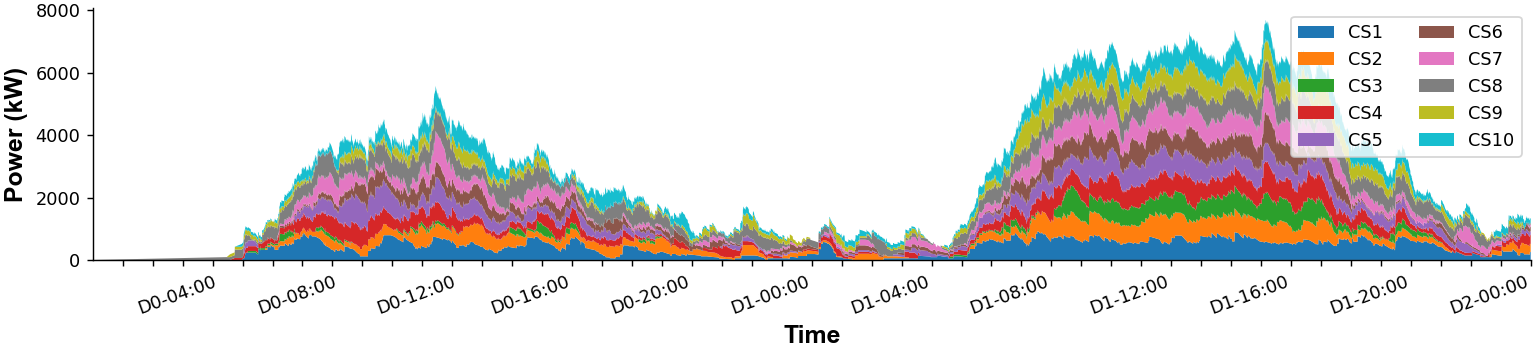}%
        \label{res_ne}}
    \hfil
    \subfloat[TAP-UE]{\includegraphics[width=3.5in]{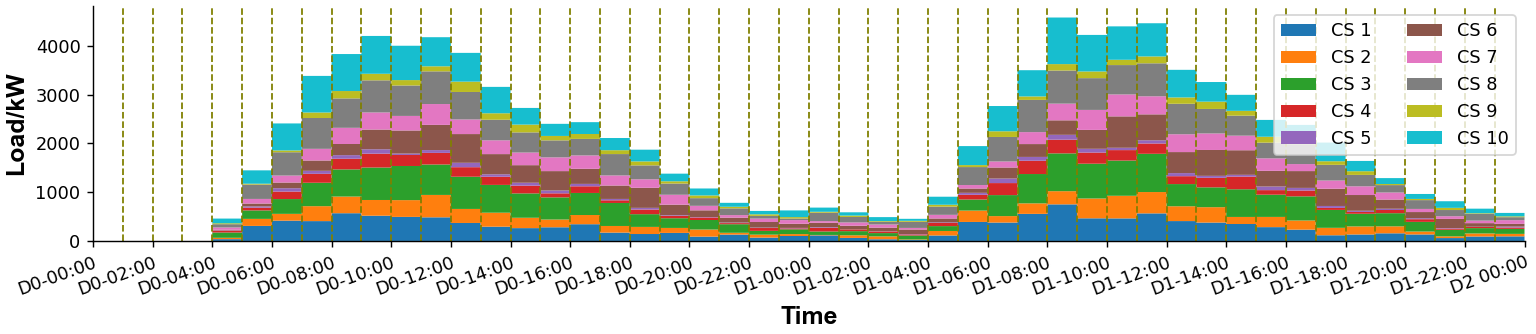}%
        \label{res_ue}}
    \caption{Comparison of the proposed platform and TAP-UE}%
    \label{res_nue_ue_comp}
\end{figure}

The TAP-UE method consumes 9218 seconds, while V2Sim consumes 165 seconds. The comparison result is shown in Figure \ref{res_nue_ue_comp}.

From Figure \ref{res_nue_ue_comp}, it can be discovered that the first-day EVCL of V2Sim and TAP-UE is close. However, the second-day EVCL varies. The maximal EVCL is a typical metric, where the TAP-UE is lower than V2Sim significantly. Such difference may be attributed to the chronological shift of SoC distribution. The average SoC at the beginning of Day 0 (the first day) is around 0.6 for both V2Sim and TAP-UE. In V2Sim, the average SoC shifts as the EVs drive and get recharged. However, in TAP-UE, the solution of each hour is independent, with a initial average SoC of 0.6 at every beginning, which may be less accurate. From the analysis above, it can be found that V2Sim is better in capturing the average SoC shifting of the EVs.

\section{Conclusion}
This paper proposes V2Sim, an open-sourced Python-based platform for performing V2G-compatible simulation of coupled MUTN and PDN. Given the topology of UTN, the number of piles and UPP of each CS, and the parameters of EVs, EVCL can be generated by simulation. Besides, parameters about EVs, buses, generators, and CSs can be recorded and plotted.

Utilizing V2Sim, the impact of several factors on the EVCL are considered. When the number of SCS piles increased, the EVCL will transfer from FCS to SCS. When the number of FCS piles increased, the PL of FCS EVCL will increase to a summit and then remain stable. When the price difference exists, the EVs will gather at lower price FCSs. When V2G is enabled at certain period, the SCS load will demonstrate a deep but narrow plunge. To accelerate the simulation procedure, multi-processing technology is employed, reduce the time consumed up to 76.6\% on a 16-core laptop. Compared to UE models, the time V2Sim used is still competitive.

The proposed SUMO-based V2G-compatible simulation program is able to consider the influence of the microscopic behavior of EVs, measure the implications of V2G on EVCL, and analyze the basic information of the PDN. V2Sim may be utilized for various coupled MUTN and PDN simulation research and other purposes. In the future, the simulation speed may be increased by employing GPU, and the V2G strategy used in V2Sim may be improved or customized. The PDN optimizer used in V2Sim is relatively simple, and more accessible interface may be added to make the data exchange between different software more convenient.


\end{document}